\documentclass[11pt]{article}

\usepackage{amsfonts}
\usepackage{appendix}
\usepackage{verbatim}

\usepackage{amsmath}
\usepackage{ulem}
\usepackage{amssymb}
\usepackage{latexsym}
\usepackage{graphicx}
\usepackage[english]{babel}
\usepackage[font={small}]{caption}
\usepackage{mathtools}
\usepackage{geometry}                
\usepackage{epstopdf}
\usepackage{tikz-cd}
\usetikzlibrary{cd}
\usepackage{amsfonts}
\usepackage{latexsym}
\usepackage{graphicx}
\usepackage[english]{babel}
\usepackage{tikz}

\begin{document}
\input epsf

\def\p{\partial}
\def\h{{1\over 2}}
\def\be{\begin{equation}}
\def\bea{\begin{eqnarray}}
\def\ee{\end{equation}}
\def\eea{\end{eqnarray}}
\def\d{\partial}
\def\la{\lambda}
\def\eps{\epsilon}
\def\b{\bigskip}
\def\m{\medskip}

\newcommand{\newsection}[1]{\section{#1} \setcounter{equation}{0}}

\def\q{\quad}

\def\h{{1\over 2}}
\def\t{\tilde}
\def\r{\rightarrow}
\def\nn{\nonumber\\}

\let\p=\partial

\newcommand\blfootnote[1]{%
  \begingroup
  \renewcommand\thefootnote{}\footnote{#1}%
  \addtocounter{footnote}{-1}%
  \endgroup
}

\begin{flushright}
\end{flushright}
\vspace{20mm}
\begin{center}
{\LARGE Lifting at higher levels in the D1D5 CFT}
\\
\vspace{18mm}
{\bf   Bin Guo$^1$\blfootnote{$^{1}$guo.1281@osu.edu} and Samir D. Mathur$^2$\blfootnote{$^{2}$mathur.16@osu.edu}
\\}
\vspace{10mm}
Department of Physics,\\ The Ohio State University,\\ Columbus,
OH 43210, USA\\ \vspace{8mm}

\vspace{8mm}
\end{center}

\vspace{4mm}

\thispagestyle{empty}
\begin{abstract}

The D1D5P system has a large set of BPS states at its orbifold point. Perturbing away from this 'free' point leads to some states joining up into long supermultiplets and lifting, while other states  remain BPS. We consider the simplest orbifold which exhibits this lift: that with $N=2$ copies of the free $c=6$ CFT. We write down the number of lifted and unlifted states implied by the index at all levels upto $6$. We work to second order in the perturbation strength $\lambda$. For levels upto $4$, we find the wavefunctions of the lifted states, their supermultiplet structure and the value of the lift. All states that are allowed to lift by the index are in fact lifted at order $O(\lambda^2)$.  We observe that the unlifted states in the untwisted sector have an  antisymmetry between the copies in the right moving Ramond ground state sector, and extend this observation to find classes of states for arbitrary $N$ that will remain unlifted to $O(\lambda^2)$. 

\vspace{3mm}

\end{abstract}
\newpage

\setcounter{page}{1}

\numberwithin{equation}{section} 

\tableofcontents

\newpage

\section{Introduction}

Black holes in string theory must be made by taking  bound states of objects -- strings and branes -- present in the theory. The D1D5P system provides a very useful example of such a construction. One finds that   the entropy and the rate of low energy emission from string states matches the expectations from gravitational thermodynamics   \cite{sv, cm, dmcompare, maldastrom}. The D1 and D5 branes form a bound state whose dynamics can be given as an effective $1+1$ dimensional conformal field theory. The momentum charge P is given by the difference in energy between the left moving and right moving excitations in this CFT.

The CFT has a `free point' which is given by a $1+1$ dimensional sigma model whose target space is an orbifold \cite{Vafa:1995bm,Dijkgraaf:1998gf,orbifold2,Larsen:1999uk,Arutyunov:1997gt,Arutyunov:1997gi,Jevicki:1998bm,David:2002wn}. The orbifold theory consists of $N$ copes of a $c=6$ CFT, joined up in different `twist sectors'.  In each twist sector the excitations are just given by free left and right moving bosons and fermions, with an overall symmetry condition to enforce the orbifold symmetry. At this orbifold point,  any state with no right moving oscillator excitations is extremal. 

The situation changes as we deform the theory away from the orbifold point. Sets of extremal states can join up into larger multiplets and lift to higher energies, leaving a smaller set of states that remain extremal.  The  count of states that remain unlifted is given by an index. This index was computed in \cite{sv} for the case where the compactification is $K3\times S^1$ and in \cite{mms} for the compactification $T^4\times S^1$.  Our interest is in finding the  actual states that are unlifted, the supermultiplet structure  for groups of states that do lift, and the value of this lift. In particular we are interested in understanding the value of the twist for sectors where typical  lifted and unlifted states arise, since this is relevant for the  physical picture of the extremal hole. States in highly twisted sectors correspond to gravity states with deep throats, while states in sectors with low twist describe shallow throats. 
(For constructions of `fuzzball' microstates dual to different CFT states see for example \cite{fuzzballs_i,fuzzballs_ii,fuzzballs_iii,fuzzballs_iv,fuzzballs_v}.)

In \cite{gn} the lift was computed, in a certain  approximation scheme,  for the situation where  most of the CFT copies are in the untwisted sector and one set is in a twisted sector. Low energy excitations of this sector can be mapped, in the gravity dual,  to strings in an $AdS_3\times S^3\times T^4$ spacetime. Apart from a small set of states in the graviton multiplet, these string states are all lifted. On the other hand we know from the index computation of \cite{mms} that if we go to sufficiently high energies and twists to reach black hole states, then a large number of states must remain {\it unlifted}: the index of \cite{mms} agrees with the Bekenstein entropy of extremal holes for large charges. It would be very interesting to understand better what properties of the highly excited states makes them remain `unlifted'. 

In this paper, we consider a CFT with $N=2$; this means that the product of the number $N_1$ of D1 branes and $N_5$ of D5 branes is $N_1 N_5\equiv N=2$.  We work to second order in the perturbation off the orbifold point. In \cite{Guo:2019ady} this problem was studied  for the lowest nontrivial amount of momentum charge $P=1$, and the pattern of lifting was found. In the present paper, we will extend the results to $P=4$ and make some observations about the general nature of the unlifted states.

\b

Our steps and results are as follows:

\b

(i) We study the constraints on lifting from the index. We find the number of states that can be lifted upto level $P=6$. We describe the structure of the long supermultiplets which relate these states.
We identify the class of states that can be lowest members of long supermultiplets.

\b

(ii) We give explicitly the wavefunctions for these lowest members of the long supermultiplets,  upto level $P=3$. (The wavefunctions for   $P=4$ is found in a similar way but the expressions are rather unwieldy, so we do not write them down in this paper.) We calculate the lift for all states to level $P=4$  at  $O(\lambda^2)$ in the perturbation. We find that at this order the lift is nonzero for all these states. Thus all states upto this level that are allowed to lift by the index are in fact lifted at order $O(\lambda^2)$.

\b

(iii) We  discuss the general nature of states that remain unlifted at $O(\lambda^2)$. First consider the case $N=2$, and states in the untwisted sector. Let the left and right sectors both be fermionic, so that the overall state of each multiwound copy is bosonic. We observe that if the right moving sector is antisymmetric in the two copies, then the state will remain unlifted at $O(\lambda^2)$. We then extend this observation to the case $N>2$, including the situation  where the component strings that are joined may have windings $k_1, k_2$ greater than unity.

\b

Before proceeding,  we note that there are many earlier works that study conformal perturbation theory, the lifting  of the states, the acquiring of anomalous dimensions, and the issue of operator mixing,   in particular in the context of the D1D5 CFT  see for example \cite{Avery:2010er,Avery:2010hs,Pakman:2009mi,Burrington:2012yq,Burrington:2014yia,Burrington:2017jhh,Carson:2016uwf}. Also, for more computations in  conformal perturbation theory in two and higher dimensional CFTs see, e.g.  \cite{kadanoff,Dijkgraaf:1987jt,Cardy:1987vr,Kutasov:1988xb,Eberle:2001jq,Gaberdiel:2008fn,Berenstein:2014cia,Berenstein:2016avf,gz,hmz,Keller:2019suk,Keller:2019yrr}.

\section{The D1D5 CFT}

In this section, we summarize some properties of the D1D5 CFT at the orbifold point and the deformation operator that we will use to perturb away from the orbifold point. For more details, see \cite{Avery:2010er,Avery:2010hs}.

Consider type IIB string theory, compactified as
\be
M_{9,1}\rightarrow M_{4,1}\times S^1\times T^4.
\label{compact}
\ee
Wrap $N_1$ D1 branes on $S^1$, and $N_5$ D5 branes on $S^1\times
T^4$. The bound state of these branes is described by a field
theory. We think of the $S^1$ as being large compared to the $T^4$, so
that at low energies we look for excitations only in the direction
$S^1$.  This low energy limit gives a conformal field theory (CFT) on
the circle $S^1$.

It has been conjectured that we can move in the moduli space of couplings in the string theory to a point called the `orbifold point' where the CFT is particularly simple. At this orbifold point the CFT is
a 1+1 dimensional sigma model. We will work in the Euclidized theory, where
the base space is a cylinder spanned by the coordinates 
\be
\tau, \sigma: ~~~0\le \sigma<2\pi, ~~~-\infty<\tau<\infty
\ee
The target space of the sigma model is the `symmetrized product' of
$N_1N_5$ copies of $T^4$,
\be
(T^4)^{N_1N_5}/S_{N_1N_5},
\ee
with each copy of $T^4$ giving 4 bosonic excitations $X^1, X^2, X^3,
X^4$. It also gives 4 fermionic excitations, which we call $\psi^1,
\psi^2, \psi^3, \psi^4$ for the left movers, and $\bar\psi^1,
\bar\psi^2,\bar\psi^3,\bar\psi^4$ for the right movers. The fermions can be
antiperiodic or periodic around the $\sigma$ circle. If they are
antiperiodic on the $S^1$ we are in the Neveu-Schwarz (NS) sector, and
if they are periodic on the $S^1$ we are in the Ramond (R)
sector. The central charge of the theory with fields
$X^i, \psi^i, ~i=1\dots 4$ is $c=6$. 
The total central charge of the entire system is thus 
\be
c=6 N_1N_5\equiv 6N
\ee

\subsection{Symmetries of the CFT}

The D1D5 CFT has $(4,4)$ supersymmetry, which means that we have
$\mathcal{N}=4$ supersymmetry in both the left and right moving
sectors.
This leads to a superconformal ${\cal N}=4$ symmetry in both
the left and right sectors, generated by operators $L_{n}, G^\pm_{\pm,r},
J^a_n$ for the left movers and $\bar L_{n}, \bar G^\pm_{\pm,r}, \bar
J^a_n$ for the right movers. The full symmetry is actually larger: it is the contracted large $\mathcal{N}=4$ superconformal symmetry \cite{mms,Sevrin:1988ew}. The algebra generators and commutators are given in Appendix~\ref{commutators}. 

Each ${\cal N} = 4$ algebra has an internal R symmetry group
$SU(2)$, so there is
a global symmetry group $SU(2)_L\times SU(2)_R$.  We denote the
quantum numbers in these two $SU(2)$ groups as
\be
SU(2)_L: ~(j, m);~~~~~~~SU(2)_R: ~ (\bar j, \bar m).
\ee
In the geometrical setting of the CFT, this symmetry arises from the
rotational symmetry in the 4 space directions of $M_{4,1}$:  we have $SO(4)_E\simeq SU(2)_L\times SU(2)_R$.
Here the subscript $E$ stands for `external', which denotes that these
rotations are in the noncompact directions.  We have another $SO(4)$ symmetry in the four directions
of the $T^4$. This symmetry we call $SO(4)_I$ (where $I$ stands for
`internal'). This symmetry is broken by the compactification of the
torus, but at the orbifold point it still provides a useful organizing
principle. We write $SO(4)_I\simeq SU(2)_1\times SU(2)_2$.
We use spinor indices $\alpha, \bar\alpha$ for $SU(2)_L$ and $SU(2)_R$
respectively. We use spinor indices $A, \dot A$ for $SU(2)_1$ and
$SU(2)_2$ respectively.

The 4 real fermions of the left sector can be grouped into complex
fermions $\psi^{\alpha A}$. The right fermions have indices $\bar{\psi}^{\bar\alpha  A}$. The bosons $X^i$ are a vector in the
$T^4$. One can decompose this vector into the $(\h, \h)$  representation of $SU(2)_1\times SU(2)_2$, which gives  scalars $X_{A\dot A}$.

\subsection{Deformation of the CFT}

The deformation of the CFT off the orbifold point is given by adding a deformation operator $D$ to the Lagrangian
\be\label{defor S}
S\r S+\lambda \int d^2 z D(z, \bar z)
\ee
where $D$ has conformal dimensions $(h, \bar h)=(1,1)$. A choice of $D$ which is a singlet under all the symmetries at the orbifold point is
\be\label{D 1/4}
D=\frac{1}{4}\epsilon^{\dot A\dot B}\epsilon_{\alpha\beta}\epsilon_{\bar\alpha \bar\beta} G^{\alpha}_{\dot A, -\h} \bar G^{\bar \alpha}_{\dot B, -\h} \sigma^{\beta \bar\beta}
\ee
where $\sigma^{\beta\bar\beta}$ is a twist operator of rank $2$ in the orbifold theory. Here $ G$ and $\bar G$ are the left and right moving supercharge operators at the orbifold point.

\section{Computation of the lift using the Gava-Narain method}

We are interested in finding states which have well defined scaling dimensions, and the values of these dimensions, as we move away from the orbifold point. We will work in the Ramond sector.  
We measure the dimensions from the Ramond ground state, which is $({c\over 24}, {c\over 24})$.  WE use the term  `level $n$' for the states  with dimensions 
\be
(h, \bar h)=(n, 0)
\label{sthree}
\ee
Let these states be labelled by  indices $a, b, \dots$, and written as $\Big|O^{(0)}_{a}\Big\rangle$ etc. 

 It turns out that while such states receive corrections at first order in $\lambda$, the dimensions get corrections only starting at $O(\lambda^2)$.   The computation involves pulling down two copies of the deformation operator $D$ from the action, and then integrating the positions of these two $D$ operators.  We first compute the matrix elements
\bea\label{Xamp}
X_{ba}(T)=\Big\langle O^{(0)}_{b}\left(\frac{T}{2}\right)\Big|\left(\int d^{2}w_{1}D(w_{1},\bar w_{1})\right)\left(\int d^{2}w_{2}D(w_{2},\bar w_{2})\right)\Big|O^{(0)}_{a}\left(-\frac{T}{2}\right)\Big\rangle
\eea
Then we compute the matrix
\bea\label{lift matrix}
E^{(2)}_{ba}=\lim_{T\rightarrow \infty}-\frac{\lambda^2}{2T}e^{E^{(0)}T}X_{ba}(T)
\eea
where $E^{(0)}$ is the energy of the states $|O^{(0)}_{a}\rangle$ at  the orbifold point.  The eigenstates of this matrix then give the linear combinations of the $|O^{(0)}_{a}\rangle$ which have definite dimensions and the eigenvalues give the lift in energy of the corresponding states. 

Such $O(\lambda^2)$ corrections were computed for some simple states in \cite{gz,hmz}. In general the computation of a correlation functions with deformation operators involves going to a covering space where the effect of the twists is undone, and one gets a correlator of operators not involving twists on this covering space. But the covering space can be a sphere in some cases, and a torus in other cases. While correlators on a sphere are easy to compute, they can be difficult to find on a torus. (A central reason for this difficulty is that the correlators on the covering space can involve spin fields. On a sphere we can remove these spin fields by spectral flows, but it is not clear how to do this on a higher genus surface.)

If we cannot explicitly compute the amplitudes (\ref{Xamp}), then how can we find the lifting? 
In \cite{gn} Gava and Narain gave a method by which amplitudes like  (\ref{Xamp}) could be written as modulus squared of amplitudes involving just {\it one} twist. Computing these one-twist amplitudes always gives a covering space that is a sphere, so the computation is straightforward. 
 
In \cite{Guo:2019pzk} this proposal of \cite{gn} was studied in detail. Let us recall the results of this study. We find
\be
\epsilon_{\dot A\dot B}\epsilon^{\bar \alpha\bar \beta}E^{(2)}_{ba}=2\lambda^2\Big\langle O^{(0)}_{b}\Big|\Big\{  \bar G^{\bar\alpha(P)}_{\dot A,0},  \bar G^{\bar \beta(P)}_{\dot B,0} \Big\}\Big|O^{(0)}_{a}\Big\rangle
\ee
We will refer to the matrix  $E_{ba}^{(2)}$ as the lifting matrix $E^{(2)}$.
The operators $\bar G^{\bar\alpha(P)}_{\dot A,0}$ are defined as
\bea\label{GN p s}
\bar G^{\bar \alpha (P)}_{\dot A,0}= \pi \mathcal P G^{+}_{\dot A,-\frac{1}{2}}\sigma^{-\bar \alpha} \mathcal P
\eea
where the  operator $\mathcal P$ is a projection operator, which projects any state to the subspace spanned by the unperturbed states $|O^{(0)}_{a}\rangle$ which have the dimensions (\ref{sthree}).

From the above relation we see that  the lifting matrix (\ref{lift matrix}) can be written using either of  the following two equivalent  expressions
\bea\label{liftmatrix 1}
E^{(2)}_{ba}=2 \lambda^2   
\Big\langle O^{(0)}_{b}\Big|    \Big\{  \bar G^{+(P)\dagger}_{+,0},  \bar G^{+(P)}_{+,0} \Big\} \Big|O^{(0)}_{a}\Big\rangle
=2 \lambda^2   
\Big\langle O^{(0)}_{b}\Big|    \Big\{  \bar G^{+(P)\dagger}_{-,0},  \bar G^{+(P)}_{-,0} \Big\} \Big|O^{(0)}_{a}\Big\rangle
\eea

Further, it was noted in  \cite{Guo:2019pzk} that the operators $\bar G^{\bar \alpha (P)}_{\dot A,0}$  give the supersymmetric structure of long multiplets. At the orbifold point the states can be grouped into short multiplets. As we deform away from the orbifold point, four of these short multiplets can join into a long multiplet and lift. The structure of this long multiplet is indicated in the following diagram:
\be\label{multiplets diagram}
\begin{tikzcd}
                                                & \phi_{+} \arrow[dr,"\bar G^{+(P)}_{-,0}"]\\
\phi \arrow[ur, "\bar G^{+(P)}_{+,0}"]  \arrow{dr}[swap]{\bar G^{+(P)}_{-,0}}& &  \phi_{+-}
\\
                                                & \phi_{-} \arrow{ur}[swap]{\bar G^{+(P)}_{+,0}}
\end{tikzcd}
~~~~~~~~~
\begin{tikzcd}
                                                & \arrow{dl} [swap]{\bar G^{-(P)}_{-,0}} \phi_{+} \\
\phi   & &  \arrow{ul}[swap]{\bar G^{-(P)}_{+,0}} \phi_{+-} \arrow{dl}{\bar G^{-(P)}_{-,0}}
\\
                                                & 
                    \arrow{ul}{\bar G^{-(P)}_{+,0}} \phi_{-} 
\end{tikzcd}
\ee
The state $\phi$ is at the bottom of this long multiplet. Note that $\phi$  is a member of a short multiplet created by operators that are not depicted in the diagram.  The operators $\bar G^{+(P)}_{+,0}$ and $\bar G^{+(P)}_{-,0}$ play the role of the two raising operators which take us to states $\phi_+, \phi_-$ which are members of two other short multiplets.  Acting with both these raising operators takes us to the short multiplet represented by the state $\phi_{+-}$.  We can move along this multiplet in the reverse direction using the lowering operators $\bar G^{-(P)}_{+,0}$ and $\bar G^{-(P)}_{-,0}$. 

Suppose we have diagonalized the matrix $E^{(2)}$ given in  eq. (\ref{liftmatrix 1}). Let  $|O^{(0)}\rangle$ be an eigenstate of this matrix. Let the corresponding eigenvalue,  which gives the lift of this operator, be called $E^{(2)}_{O}$. Then from (\ref{liftmatrix 1}) we find that $E^{(2)}_O$ can be written as a sum of modulus-squared terms
\be\label{lifting norm}
E^{(2)}_O=2 \lambda^2   
\Big(\Big|\bar G^{+(P)}_{+,0} |O^{(0)}\rangle\Big|^2+\Big|\bar G^{-(P)}_{-,0} |O^{(0)}\rangle\Big|^2\Big)
=2 \lambda^2   
\Big(\Big|\bar G^{+(P)}_{-,0} |O^{(0)}\rangle\Big|^2+\Big|\bar G^{-(P)}_{+,0} |O^{(0)}\rangle\Big|^2\Big)
\ee

In the long multiplet described in (\ref{multiplets diagram}) each of the four states $\phi$, $\phi_{+}$, $\phi_{-}$ and $\phi_{+-}$ have the following property:  if it can be raised by $\bar G^{+(P)}_{+,0}$, then it will be annihilated by the  $\bar G^{-(P)}_{-,0}$; conversely, if it can be lowered by $\bar G^{-(P)}_{-,0}$ then it will be annihilated by  the $\bar G^{+(P)}_{+,0}$.  A similar statement holds for the raising operators $\bar G^{+(P)}_{-,0}$ and the lowering operators $\bar G^{-(P)}_{+,0}$. Thus in each of the two expressions in (\ref{lifting norm}), only one of the two terms is nonzero. 

In summary, one can get the value of the lift and the corresponding eigenstates by diagonalizing the lifting matrix $E^{(2)}$ (\ref{liftmatrix 1}). 
Four short multiplets join into a long multiplet as shown in eq. (\ref{multiplets diagram}) and the lifting can be calculated from (\ref{lifting norm}).

In the cases that we will encounter below, there is a unique state $|O^{(0)}_{a}\rangle$ with the relevant quantum numbers, so we will not have to diagonalize a matrix $E^{(2)}_{ba}$ to first find the eigenvectors $|O^{(0)}\rangle$.

\section{The character decomposition}

The CFT has a left and a right moving superconformal symmetry, with each of these described by the contracted large $\mathcal{N}=4$ superconformal algebra \cite{mms,Sevrin:1988ew}. The algebra generators and commutation relations are given in Appendix~\ref{commutators}.

These symmetries remain true for all values of the coupling. Thus states related by these symmetries will have the same lift $E^{(2)}$. We would like to group states into multiplets that are related by these symmetries, so that we can reduce the number of independent lifting computations that we have to perform. In this section, we will count the number of multiplets by using the character decompostion (see the related works \cite{b,bh}). The main result of this section is given in Table\,\ref{table chara decomp} below.

In subsection \ref{pf}, we recall the partition function of the orbifold theory. 
In subsection \ref{sec chara}, we count the number of multiplets by writing the partition function in terms of characters. In subsection \ref{index}, we consider the constraints on lifting that arise from considering the  index.

\subsection{The partition function}\label{pf}

In this subsection we recall the partition function of the orbifold  CFT. 
The partition function for a single $c=6$ copy of the CFT  is defined as
\be\label{pf single copy}
Z={\rm Tr} (-1)^{2J^{3}_{0}-2\bar J^{3}_{0}}q^{L_{0}-c/24}\bar q^{\bar L_{0}-c/24} y^{2J^{3}_{0}} \bar y^{2 \bar J^{3}_{0}}\equiv\sum_{h,\bar h,j_3,\bar j_3} c(h,\bar h,j_{3},\bar j_{3})q^{h}\bar q^{\bar h} y^{2j_3} \bar y^{2\bar j_3}
\ee
We work with the case where  the 4-manifold is $T^4$. There are $U(1)$ charges that arise from momentum and winding charges around this $T^4$, but we work in the sector where all such charges have been taken to be zero. For this choice, one finds
\be
Z(T^4)=\left(\frac{\theta_{1}}{\eta}\right)^{2}\frac{1}{\eta^4} \overline{\left(\frac{\theta_{1}}{\eta}\right)^{2}\frac{1}{\eta^4}}
\ee
where
\bea
\theta_{1}&=&i(y^{1/2}-y^{-1/2})q^{1/8}\prod_{n=1}^{\infty}(1-q^n)(1-y q^n)(1-y^{-1}q^n)\nn
\eta&=&q^{1/24}\prod_{n=1}^{\infty}(1-q^{n})
\eea
Using the above we can find the partition function for the case where the target space of the $1+1$ dimensional CFT is the symmetric product
 $Sym^{N}(T^4)$.  The partition function for a symmetric product target space
 $Z(Sym^{N}(X))$ is given by 
\be\label{partition function}
\mathcal Z (p,q,\bar q, y, \bar y)=\sum^{\infty}_{N=0}p^{N}Z(Sym^{N}(X))= \prod_{n=1}^{\infty} \prod'_{h,\bar h, j_3,\bar j_3} \frac{1}{(1-p^{n}q^{h/n}\bar q^{\bar h/n} y^{2j_3} \bar y^{2\bar j_3})^{c(h,\bar h,j_{3},\bar j_{3})}}
\ee
where $\prod'_{h,\bar h, j_3,\bar j_3}$ is restricted so that $(h-\bar h)/n$ is an integer. The $c(h,\bar h,j_{3},\bar j_{3})$ are the degeneracies appearing in the partition function for the CFT with a single copy of the space $X$. 
We write 
\be\label{expand p f}
\mathcal Z (p,q,\bar q, y, \bar y)=\sum_{N,h,\bar h,j_{3},\bar j_{3}} c(N,h,\bar h,j_{3},\bar j_{3})p^{N} q^{h}\bar q^{\bar h} y^{2j_3} \bar y^{2\bar j_3}
\ee
This yields the degeneracies $c(N,h,\bar h,j_{3},\bar j_{3})$ of the states with total winding of the effective string $N$ and  quantum number $h,\bar h, j_{3},\bar j_{3}$.

\subsection{The character decomposition}\label{sec chara}

We now write the partition function in terms of the characters of the contracted large $\mathcal{N}=4$ superconformal symmetry.
We let the total winding number of the effective string be $N=2$, and consider states with  $\bar h=0$. From (\ref{partition function}) and (\ref{expand p f}), the partition function for this case  is
\be\label{expand p f 2}
Z(N=2;\bar h=0)=\sum_{h,j_{3},\bar j_{3}} c(N=2,h,\bar h=0,j_{3},\bar j_{3}) q^{h} y^{2j_3} \bar y^{2\bar j_3}
\ee

There are two different twist sectors for $N=2$: the case of two singly wound copies which we call $N=(1,1)$, and the case of a single doubly wound copy which we call $N=(2)$. The expression (\ref{expand p f 2}) includes the contribution from both these twist sectors. We are however interested in obtaining the contribution separately from these two different twist sectors. It turns out that with a little effort we can separate the two contributions in the expression (\ref{expand p f 2}). This is done as follows:

\b

N=(1,1) sector: In (\ref{partition function}), restrict the  product to terms  with $n=1$  and collect all the terms with dependence $p^2$.

\b

N=(2) sector: In (\ref{partition function}), restrict the  product to terms  with $n=2$  and collect all the terms with dependence $p^2$.

\b

In this way, we can separate the two contributions
\be
Z(N=2;\bar h=0)=Z(N=(1,1); \bar h=0)+Z(N=(2); \bar h=0)
\ee
For each of these two sectors, we find  the following character decomposition
\be\label{chara decomp}
Z(q, y, \bar y)=\sum_{\substack{j=1/2,1 \\ \bar j_{3}=-1,-1/2,0}}c^s_{j;\bar j_{3}}\chi^{s}_{j}(q,y)\bar \chi_{\bar j_{3}}( \bar y)+\sum_{\substack{j=1 \\ \bar j_{3}=-1,-1/2,0}}c^l_{j,h;\bar j_{3}}\chi^{l}_{j,h}(q,y)\bar \chi_{\bar j_{3}}( \bar y)
\ee
where $\chi^{s}_{j}$ and $\chi^{l}_{j,h}$ are given in Appendix\,\ref{character}. 
They are characters of a left moving contracted large $\mathcal N=4$ algebra. 
The $\chi^{s}_{j}$ is the character for a short representation in which the primary has dimension $h=0$. 
The $\chi^{l}_{j,h}$ is the character for a long representation in which the primary has dimension $h>0$. 
From Appendix\,\ref{character} for $N=2$, we see that the possible values of $j$ for a short representation are $j=1/2,1$ and for a long representation is $j=1$.

For the character $\bar \chi_{\bar j_{3}}$ of the right mover, we consider the sub-algebra formed by $\bar d^{\bar\alpha A}_0$.
The lowest weight states 
are defined by
\be\label{right lowest}
\bar d^{- A}_{0}| \phi\rangle=0
\ee
with  charge 
\be
\bar J^{3}_{0}| \phi\rangle=\bar j_{3} | \phi\rangle
\ee
There are two fermionic raising operators $\bar d^{+ \pm}_{0}$. Application of these operators once to a lowest weight state, which has charge $\bar j_3$, gives two states with charges $\bar j_{3}+1/2$. Applying again gives a state with charge $\bar j_{3}+1$
\be
\bar j_{3} \rightarrow 2(\bar j_{3}+1/2) \rightarrow \bar j_{3}+1 
\ee
The character is defined by the trace over the irreducible representation. We find
\be
\bar \chi_{\bar j_{3}}(\bar y)={\rm Tr} (-1)^{-2\bar J^{3}_{0}} \bar y^{2\bar J^{3}_{0}}
=-(-\bar y)^{2\bar j_{3}+1}(\bar y^{1/2}-\bar y^{-1/2})^2
\ee
The possible values of $\bar j_{3}$ for $N=2$ are $\bar j_{3}=-1,-1/2,0$. The results of the character decomposition (\ref{chara decomp}) of $N=2$ are

\b

(i) For the coefficients $c^s_{j;\bar j_{3}}$ in (\ref{chara decomp}), which tell us the numbers of primaries with dimension $h=0$ and with charges $j$ and $\bar j_3$, the results are
\bea
(1,1)~&:& 4\chi^{s}_{j=1/2}\bar \chi_{\bar j_{3}=-1/2}+\chi^{s}_{j=1}\bar \chi_{\bar j_{3}=-1}+\chi^{s}_{j=1}\bar \chi_{\bar j_{3}=0}\nn
(2)~~&:& \chi^{s}_{j=1/2}\bar \chi_{\bar j_{3}=-1/2}
\eea

\b

(ii) For the coefficients $c^l_{j,h;\bar j_{3}}$, which tell us the numbers of primaries with dimension $h>0$ and with charges $j$ and $\bar j_3$, the results are shown in table \ref{table chara decomp}.

\begin{table}[ht!]
\caption{The character decomposition at the orbifold point}
\begin{center}
\begin{tabular}{ |c|c|c|c|c|c|c| } 
 \hline
level & sector & $\chi^{l}_{j=1,h}\bar \chi_{\bar j_{3}=-1}$  & $\chi^{l}_{j=1,h}\bar \chi_{\bar j_{3}=-1/2}$ & $\chi^{l}_{j=1,h}\bar \chi_{\bar j_{3}=0}$& sector & $\chi^{l}_{j=1,h}\bar \chi_{\bar j_{3}=-1/2}$ \\
\hline
 & & $\phi$  & unlifted & $\phi_{+-}$ &    & $\phi_{+}$,$\phi_{-}$,unlifted \\ 
 \hline
$h=1$ &(1,1)& 3 & 0 & 3 & (2)  & 6 \\ 
 \hline
$h=2$ &(1,1)& 1 & 16 & 1 & (2)  & 28 \\ 
 \hline
$h=3$ &(1,1)& 18 & 8 & 18 & (2)  & 98 \\ 
 \hline
$h=4$ &(1,1)& 15 & 72 & 15 & (2)  & 282 \\ 
 \hline
$h=5$ &(1,1)& 68 & 80 & 68 & (2)  & 728 \\
\hline
$h=6$ &(1,1)& 89 & 264 & 89 & (2)  & 1734 \\ 
 \hline
\end{tabular}
\end{center}
\label{table chara decomp}
\end{table}

\subsection{The constraints on lifting from the index}\label{index}

Let us first recall how the index is computed. Consider the  exact supercharge operators $\bar G^{+}_{\dot A}$, $\dot A=+,-$ of the perturbed CFT. These operators are the ones that join four short multiplets  into a long multiplet. Let us see the structure of the set of states that will join into a long multiplet.
Each of the operators $\bar G^{+}_{\dot A}$ increases the $SU(2)_R$ charge by 1/2 and does not change the $SU(2)_L$ charge.
Thus the four short multiplets joining into a long multiplet must have the same left moving character but their right moving characters  will be as follows:
\be\label{right chi long}
\bar \chi_{\bar j_{3}}~~~~~~2\bar \chi_{\bar j_{3}+1/2}~~~~~\bar \chi_{\bar j_{3}+1}
\ee
Whenever we can group states in the manner indicated by such a set of characters, then we find a set of states that have the charges to join into a long multiplet. We therefore exclude such sets of states from the index. If there are states left over that {\it cannot} group into a set with characters (\ref{right chi long}), then we count those states in the index, since they cannot possibly join into a long multiplet and lift.

Now let us look at table \ref{table chara decomp}. In the row  $h=1$, it is possible to join  
$3\chi^{l}_{j=1,h}\bar \chi_{\bar j_{3}=-1}$ and $3\chi^{l}_{j=1,h}\bar \chi_{\bar j_{3}=0}$ in the (1,1) sector and $6 \chi^{l}_{j=1,h}\bar \chi_{\bar j_{3}=-1/2}$ in the (2) sector into 3 long multiplets.
In second order perturbation theory, they indeed join into 3 long multiplets \cite{Guo:2019ady}.
In the row  $h=2$, it is possible to join  $\chi^{l}_{j=1,h}\bar \chi_{\bar j_{3}=-1}$ and $\chi^{l}_{j=1,h}\bar \chi_{\bar j_{3}=0}$ in (1,1) sector and $2\chi^{l}_{j=1,h}\bar \chi_{\bar j_{3}=-1/2}$ into a long multiplet. The $2$ of $\chi^{l}_{j=1,h}\bar \chi_{\bar j_{3}=-1/2}$ can come from the $16$ in the (1,1) sector or the $28$ in the (2) sector. In  second order perturbation theory, it should come from the (2) sector because the operators (\ref{GN p s}) joining short multiplets into long multiplets change the twist sector. 

In general, in second order perturbation theory for any level $h>0$, the operators (\ref{GN p s}) join $\chi^{l}_{j=1,h}\bar \chi_{\bar j_{3}=-1}$ and $\chi^{l}_{j=1,h}\bar \chi_{\bar j_{3}=0}$ in the (1,1) sector and $2\chi^{l}_{j=1,h}\bar \chi_{\bar j_{3}=-1/2}$ in the (2) sector into a long multiplet. In  Table \ref{table chara decomp}, the number in the column $\chi^{l}_{j=1,h}\bar \chi_{\bar j_{3}=-1/2}$ in the (2) sector grows faster than twice of the number in the column $\chi^{l}_{j=1,h}\bar \chi_{\bar j_{3}=-1}$ or $\chi^{l}_{j=1,h}\bar \chi_{\bar j_{3}=0}$ in the (1,1) sector. Thus it is possible that all the states in the columns $\chi^{l}_{j=1,h}\bar \chi_{\bar j_{3}=-1}$ and $\chi^{l}_{j=1,h}\bar \chi_{\bar j_{3}=0}$ in the (1,1) sector will be lifted by pairing with states in the (2) sector. In the following, we will find that states in these two columns with dimension up to $h=4$ are indeed lifted.

\section{The method for finding the lifted primaries}

The primaries and their descendents have same lift. Thus we can focus on the primaries, which are counted in table \ref{table chara decomp}. To find all the lifted primaries, we need to construct the primaries in the $\chi^{l}_{j=1,h}\bar \chi_{\bar j_{3}=-1}$ column explicitly. Then applying the two raising operators $\bar G^{+ (P)}_{\pm ,0}$ gives all the four lifted primaries in a long multiplet as in (\ref{multiplets diagram}).

In subsection \ref{primaries right}, we will construct the right movers of all the primaries explicitly.
In subsection \ref{primaries left}, we will construct the left movers of the primaries in the $\chi^{l}_{j=1,h}\bar \chi_{\bar j_{3}=-1}$ column up to level $h=4$.

\subsection{Right movers}\label{primaries right}
The right movers of primaries satisfying (\ref{right lowest}) with total winding $N=2$ were found in \cite{Guo:2019ady}. There are two classes:

\b

(i) The following set of right moving states  can join into a long multiplet and  lift at second order in perturbation theory
\bea\label{right mover}
(1,1)~~\bar \chi_{\bar j_{3}=-1}~:~~~~~~~~~~~~
|\phi^{R}\rangle&=&|\bar 0^{-}_{R}\rangle|\bar 0^{-}_{R}\rangle\nn
(2)~~\bar \chi_{\bar j_{3}=-1/2}~:~~
|\phi^{R}_{+}\rangle=|\phi^{R}_{-}\rangle&=&|\bar 0^{2-}_{R}\rangle\nn
(1,1)~~\bar \chi_{\bar j_{3}=0}~~~:~~~~~~~~~~
|\phi^{R}_{+-}\rangle&=&\frac{1}{2}(\bar d_0^{++(1)}-\bar d_0^{++(2)})(\bar d_0^{+-(1)}-\bar d_0^{+-(2)})|\bar 0^{-}_{R}\rangle|\bar 0^{-}_{R}\rangle
\eea

\b

(ii) The following right moving state cannot join with other states into a long multiplet and thus states of this type cannot lift at second order in perturbation theory
\be\label{r mover zero lifting}
(1,1)~~\bar \chi_{\bar j_{3}=-1/2}~~~:~~~~~~~~~~(\bar d_0^{+A(1)}-\bar d_0^{+A(2)})|\bar 0^{-}_{R}\rangle|\bar 0^{-}_{R}\rangle
\ee

\b
The first and third right movers in (\ref{right mover}) and the right mover in (\ref{r mover zero lifting}) are the three right moving primaries listed for  the (1,1) sector in table \ref{table chara decomp}. The second right mover in (\ref{right mover}) is the right mover of the primary in the (2) sector in table \ref{table chara decomp}.

The right mover of the operator $\bar G^{\bar \alpha (P)}_{\dot A,0}$ is
$\mathcal P \bar \sigma^{\bar \alpha}$.
Thus the long multiplet structure (\ref{sthree}) for the right movers is as given in the following diagram:

\be\label{multiplets diagram right}
\begin{tikzcd}
                                                & \phi^R_{+} \arrow[dr,"\mathcal P\bar \sigma^{+}"]\\
\phi^R \arrow[ur, "\mathcal P\bar \sigma^{+}"]  \arrow{dr}[swap]{\mathcal P\bar \sigma^{+}}& &  \phi^R_{+-}
\\
                                                & \phi^R_{-} \arrow{ur}[swap]{\mathcal P\bar \sigma^{+}}
\end{tikzcd}
~~~~~~~~~
\begin{tikzcd}
                                                & \arrow{dl} [swap]{\mathcal P\bar \sigma^{-}} \phi^R_{+} \\
\phi^R   & &  \arrow{ul}[swap]{\mathcal P\bar \sigma^{-}} \phi^R_{+-} \arrow{dl}{\mathcal P\bar \sigma^{-}}
\\
                                                & 
                    \arrow{ul}{\mathcal P\bar \sigma^{-}} \phi^R_{-} 
\end{tikzcd}
\ee

The states in class (i) satisfy the properties in (\ref{multiplets diagram right}). Thus they can join into a long multiplet and have nonzero lifting. 
The states in class (ii) are annihilated by $\mathcal P \bar \sigma^{\bar \alpha}$. Thus states who have right mover (\ref{r mover zero lifting}) have zero lift at $O(\lambda^2)$.

\subsection{Left movers}\label{primaries left}

In this subsection, we will explain how to find the left movers of primaries in the $\chi^{l}_{j=1,h}\bar \chi_{\bar j_{3}=-1}$ column in Table\,\ref{table chara decomp}.

In the $(1,1)$ sector, the global modes (\ref{global modes app}) are defined as
\be
O^{(g)}_{n}=O^{(1)}_n+O^{(2)}_n
\ee
Thus these modes are applied symmetrically to the two copies of the CFT. The primaries $\phi$ with dimensions $h>0$ of the contracted large $\mathcal N=4$ superconformal algebra are defined by
\bea\label{def primary 1}
L^{(g)}_{n}|\phi\rangle= G^{+(g)}_{\dot A,n}|\phi\rangle= J^{3(g)}_{n}|\phi\rangle=J^{+(g)}_{n}|\phi\rangle=0~~~~~~~n>0\nn
G^{-(g)}_{\dot A,n}|\phi\rangle=J^{-(g)}_{n}|\phi\rangle=0~~~~~~~n\geq 0
\eea
and
\be\label{def primary 2}
\alpha^{(g)}_{A\dot A,n}|\phi\rangle= d^{\alpha A(g)}_{n}|\phi\rangle=0~~~~~~~n>0
\ee
From the algebra in Appendix \ref{commutators},  all the conditions in eq.\,(\ref{def primary 1}) follow from the conditions 
\bea\label{primary check}
J^{-(g)}_{0}|\phi\rangle= J^{+(g)}_{1}|\phi\rangle=G^{-(g)}_{\dot A,0}|\phi\rangle=0
\eea
To find all states satisfying the relations (\ref{def primary 2}), we define operators that are  antisymmetric  between the two copies:
\be\label{def anti}
O^{(\mathcal{A})}_{n}=O^{(1)}_n-O^{(2)}_n
\ee
From the algebra in Appendix \ref{commutators}, one finds
\bea
[\alpha^{(g)}_{A\dot A,n},\alpha^{(\mathcal A)}_{B\dot B,m} ]=0,~~~~~~~~
\{d^{\alpha A(g)}_{n},d^{\beta B(\mathcal A)}_{m}\}=0
\eea
Thus the states satisfying  eq.\,(\ref{def primary 2}) are the states built by acting with the antisymmetric operators $\alpha^{(\mathcal A)}_{A\dot A,n}$ and $d^{\alpha A(\mathcal A)}_{n}$ on the Ramond ground state $|0^-_R\rangle|0^-_R\rangle$.

Therefore, to find the primaries in the $\chi^{l}_{j=1,h}\bar \chi_{\bar j_{3}=-1}$ column, we can first build states using the antisymmetric operators $\alpha^{(\mathcal A)}_{A\dot A,n}$ and $d^{\alpha A(\mathcal A)}_{n}$. Then on this space of states we find the solutions of eq.\,(\ref{primary check}). This procedure gives the required primaries. The results are described  in section \ref{sec lift} below.

\section{Long multiplets and lifting}\label{sec lift}

In this section, we will find the long multiplets and their lift up to level-4.
For each level, we will first construct the primaries in the $\chi^{l}_{j=1,h}\bar \chi_{\bar j_{3}=-1}$ column. Then we will apply the two raising operators $\bar G^{+ (P)}_{\pm ,0}$ to get all the four primaries in a long multiplet. Because the four primaries in a long multiplet have the same lift, we will only calculate the lift of the primaries in the $\chi^{l}_{j=1,h}\bar \chi_{\bar j_{3}=-1}$ column. For these states, the lift (\ref{lifting norm}) becomes
\be\label{lifting phi}
E^{(2)}_{\phi}=2 \lambda^2   
\Big|\bar G^{+(P)}_{+,0} |\phi\rangle\Big|^2
=2 \lambda^2   
\Big|\bar G^{+(P)}_{-,0} |\phi\rangle\Big|^2
\ee

In the following, we will organize primaries into $A$ and $\dot A$ charge multiplets, which will be labeled by $(j_{A},j_{\dot A})$. We will only write the result of the lowest weight states explicitly for each $A$ and $\dot A$ charge multiplet.

We find that up to level-4 all states that can lift do actually lift at second order. The lifts are listed in  table \ref{table lift}.
\begin{table}[ht!]
\caption{The energy lift at second order}
\begin{center}\label{table lift}
\begin{tabular}{ |c|c|c|c|c|c|c| } 
 \hline
level & number of long multiplets  & $(j_{A},j_{\dot A})$ & $E^{(2)}/\pi^2 \lambda^2$ \\
 \hline
$h=1$ & 3  & $(1,0)$ & 1  \\
 \hline
$h=2$ & 1  & $(0,0)$ & 15/8  \\
 \hline
$h=3$ & 18 &$(2,1)$, $(1,0)$  & 3, 64/39 \\
 \hline
$h=4$ & 15 &$(1,1)$, $(2,0)$, $(0,0)$ & 3, 5/2, 2695/1024 \\
 \hline
\end{tabular}
\end{center}
\label{table lift}
\end{table}

\subsection{Level-1}

The 3 primaries in the $\chi^{l}_{j=1,h}\bar \chi_{\bar j_{3}=-1}$ column can be organized into a $A$ charge triplet and $\dot A$ charge singlet; we label this as $(1,0)$. 

For the lowest weight state of the $(1,0)$ multiplet, the normalized left movers of the long multiplet are
\bea\label{left movers}
|\phi^{L}\rangle=|\phi^{L}_{+-}\rangle
&=&\frac{1}{2}d_{-1}^{--(\mathcal A)}d_0^{+-(\mathcal A)}|0^{-}_{R}\rangle|0^{-}_{R}\rangle\nn
|\phi^{L}_{+}\rangle&=&\frac{1}{\sqrt{2}} d^{--}_{-1/2}\alpha_{++,-1/2}|0^{2-}_{R}\rangle\nn
|\phi^{L}_{-}\rangle&=&\frac{1}{\sqrt{2}} d^{--}_{-1/2}\alpha_{+-,-1/2}|0^{2-}_{R}\rangle
\eea
while the normalized right movers are in eq.\,(\ref{right mover}).
Applying (\ref{lifting phi}) to the primary $\phi$, one finds the lift
\bea
E^{(2)}_{1,(1,0)}= \pi^2 \lambda^2
\eea

\subsection{Level-2}

The one primary in the $\chi^{l}_{j=1,h}\bar \chi_{\bar j_{3}=-1}$ column is a $A$ charge singlet and $\dot A$ charge singlet; thus we have the representation  $(0,0)$.

The normalized left movers of the long multiplet are
\bea
|\phi^{L}\rangle=|\phi^{L}_{+-}\rangle
&=&\frac{1}{2\sqrt{10}}\Big[d^{-+(\mathcal A)}_{-1}d^{--(\mathcal A)}_{-1}d^{+-(\mathcal A)}_{0}d^{++(\mathcal A)}_{0}\nn
&&~~~~~~~~+d^{-+(\mathcal A)}_{-1}d^{+-(\mathcal A)}_{-1}+d^{++(\mathcal A)}_{-1}d^{--(\mathcal A)}_{-1}
-d^{-+(\mathcal A)}_{-2}d^{+-(\mathcal A)}_{0}+d^{--(\mathcal A)}_{-2}d^{++(\mathcal A)}_{0}\nn
&&~~~~~~~~-\alpha^{(\mathcal A)}_{++,-1}\alpha^{(\mathcal A)}_{--,-1}+\alpha^{(\mathcal A)}_{-+,-1}\alpha^{(\mathcal A)}_{+-,-1}\Big]|0^{-}_{R}\rangle|0^{-}_{R}\rangle\nn
|\phi^{L}_{+}\rangle&=&\frac{1}{4\sqrt{6}}\sum_{A=+,-}\Big[\alpha_{A+,-3/2}d^{-A}_{-1/2}
-3\alpha_{A+,-1/2}d^{-A}_{-3/2}\nn
&&\hspace{2cm}-\frac{3}{2}\alpha_{A+,-1/2}d^{+A}_{-1/2}d^{-+}_{-1/2}d^{--}_{-1/2}\nn
&&\hspace{1cm}+(\alpha_{+-,-1/2}\alpha_{-+,-1/2}-\alpha_{++,-1/2}\alpha_{--,-1/2})\alpha_{A+,-1/2}d^{-A}_{-1/2}
\Big]|0^{2-}_{R}\rangle\nn
|\phi^{L}_{-}\rangle&=&\frac{1}{4\sqrt{6}}\sum_{A=+,-}\Big[\alpha_{A-,-3/2}d^{-A}_{-1/2}
-3\alpha_{A-,-1/2}d^{-A}_{-3/2}\nn
&&\hspace{2cm}-\frac{3}{2}\alpha_{A-,-1/2}d^{+A}_{-1/2}d^{-+}_{-1/2}d^{--}_{-1/2}\nn
&&\hspace{1cm}+(\alpha_{+-,-1/2}\alpha_{-+,-1/2}-\alpha_{++,-1/2}\alpha_{--,-1/2})\alpha_{A-,-1/2}d^{-A}_{-1/2}
\Big]|0^{2-}_{R}\rangle
\eea
Applying (\ref{lifting phi}) to the primary $\phi$, one finds the lift
\bea
E^{(2)}_{2,(0,0)}=\frac{15}{8}\pi^2\lambda^2
\eea

\subsection{Level-3}

The 18 primaries in the $\chi^{l}_{j=1,h}\bar \chi_{\bar j_{3}=-1}$ column can be organized into a A charge quintet and $\dot A$ charge triplet and a $A$ charge triplet and $\dot A$ charge singlet; thus the representations are $(2,1)$ and $(1,0)$.

For the lowest weight state of the $(2,1)$ multiplet, the normalized left movers of the long multiplet are
\bea
|\phi^{L}\rangle=|\phi^{L}_{+-}\rangle
&=&\frac{1}{4}\alpha^{(\mathcal A)}_{+-,-1}\alpha^{(\mathcal A)}_{+-,-1}d^{--(\mathcal A)}_{-1}d^{+-(\mathcal A)}_{0}|0^{-}_{R}\rangle|0^{-}_{R}\rangle
\eea
\bea
|\phi^{L}_{+}\rangle&=&\frac{1}{2\sqrt{6}}\Big[
\alpha_{+-,-3/2}\alpha_{++,-1/2}\alpha_{+-,-1/2}d^{--}_{-1/2}
-\frac{1}{2}\alpha_{++,-3/2}\alpha_{+-,-1/2}\alpha_{+-,-1/2}d^{--}_{-1/2}\nn
&&\hspace{1cm}+\alpha_{+-,-1/2}d^{--}_{-3/2}d^{+-}_{-1/2}d^{--}_{-1/2}
+\frac{1}{2}\alpha_{++,-1/2}\alpha_{+-,-1/2}\alpha_{+-,-1/2}d^{--}_{-3/2}\nn
&&\hspace{1cm}-\frac{1}{2}\alpha_{++,-1/2}\alpha_{+-,-1/2}\alpha_{+-,-1/2}\alpha_{+-,-1/2}\alpha_{-+,-1/2}d^{--}_{-1/2}\nn
&&\hspace{1cm}+\frac{1}{2}\alpha_{++,-1/2}\alpha_{++,-1/2}\alpha_{+-,-1/2}\alpha_{+-,-1/2}\alpha_{--,-1/2}d^{--}_{-1/2}\nn
&&\hspace{1cm}+\frac{1}{4}\alpha_{++,-1/2}\alpha_{+-,-1/2}\alpha_{+-,-1/2}d^{+-}_{-1/2}d^{-+}_{-1/2}d^{--}_{-1/2}
\Big]|0^{2-}_{R}\rangle
\eea

\bea
|\phi^{L}_{-}\rangle&=&\frac{1}{2\sqrt{6}}\Big[\frac{1}{2}
\alpha_{+-,-3/2}\alpha_{+-,-1/2}\alpha_{+-,-1/2}d^{--}_{-1/2}
+\frac{1}{2}\alpha_{+-,-1/2}\alpha_{+-,-1/2}\alpha_{+-,-1/2}d^{--}_{-3/2}\nn
&&\hspace{1cm}-\frac{1}{2}\alpha_{+-,-1/2}\alpha_{+-,-1/2}\alpha_{+-,-1/2}\alpha_{+-,-1/2}\alpha_{-+,-1/2}d^{--}_{-1/2}\nn
&&\hspace{1cm}+\frac{1}{2}\alpha_{++,-1/2}\alpha_{+-,-1/2}\alpha_{+-,-1/2}\alpha_{+-,-1/2}\alpha_{--,-1/2}d^{--}_{-1/2}\nn
&&\hspace{1cm}+\frac{1}{4}\alpha_{+-,-1/2}\alpha_{+-,-1/2}\alpha_{+-,-1/2}d^{+-}_{-1/2}d^{-+}_{-1/2}d^{--}_{-1/2}
\Big]|0^{2-}_{R}\rangle
\eea
Applying (\ref{lifting phi}) to the primary $\phi$, one finds the lift
\bea
E^{(2)}_{3,(2,1)}=3\pi^2\lambda^2
\eea

For the lowest weight state of the $(1,0)$ multiplet, the normalized left movers of the long multiplet are
\bea
|\phi^{L}\rangle=|\phi^{L}_{+-}\rangle
&=&\frac{1}{2\sqrt{46}}\Big[d^{--(\mathcal A)}_{-3}d^{+-(\mathcal A)}_0-2d^{--(\mathcal A)}_{-2}d^{+-(\mathcal A)}_{-1}-d^{+-(\mathcal A)}_{-2}d^{--(\mathcal A)}_{-1}\nn
&&~~~~~~~+\frac{3}{2}d^{--(\mathcal A)}_{-2}d^{--(\mathcal A)}_{-1}d^{++(\mathcal A)}_{0}d^{+-(\mathcal A)}_{0}
-\frac{3}{2}d^{+-(\mathcal A)}_{-1}d^{--(\mathcal A)}_{-1}d^{-+(\mathcal A)}_{-1}d^{+-(\mathcal A)}_{0}\nn
&&~~~~~~~+\frac{3}{2}(\alpha^{(\mathcal A)}_{++,-1}\alpha^{(\mathcal A)}_{--,-1}-\alpha^{(A)}_{+-,-1}\alpha^{(\mathcal A)}_{-+,-1})d^{--(\mathcal A)}_{-1}d^{+-(\mathcal A)}_{0}\nn
&&~~~~~~~+\alpha^{(\mathcal A)}_{++,-2}\alpha^{(\mathcal A)}_{+-,-1}-\alpha^{(\mathcal A)}_{+-,-2}\alpha^{(\mathcal A)}_{++,-1}\Big]|0^{-}_{R}\rangle|0^{-}_{R}\rangle
\eea

\bea
|\phi^{L}_{+}\rangle&=&\frac{1}{\sqrt{897}}\Big[\frac{15}{16}\alpha_{++,-5/2}d^{--}_{-1/2}
-\frac{29}{8}\alpha_{++,-3/2}d^{--}_{-3/2}+\frac{75}{16}\alpha_{++,-1/2}d^{--}_{-5/2}\nn
&&\hspace{1cm}+\frac{27}{16}\alpha_{--,-3/2}\alpha_{++,-1/2}\alpha_{++,-1/2}d^{--}_{-1/2}-\frac{27}{16}\alpha_{-+,-3/2}\alpha_{++,-1/2}\alpha_{+-,-1/2}d^{--}_{-1/2}\nn
&&\hspace{1cm}-\frac{43}{16}\alpha_{+-,-3/2}\alpha_{++,-1/2}\alpha_{-+,-1/2}d^{--}_{-1/2}-\alpha_{+-,-3/2}\alpha_{++,-1/2}\alpha_{++,-1/2}d^{-+}_{-1/2}\nn
&&\hspace{1cm}+\frac{21}{8}\alpha_{++,-3/2}\alpha_{+-,-1/2}\alpha_{-+,-1/2}d^{--}_{-1/2}+\frac{1}{16}\alpha_{++,-3/2}\alpha_{++,-1/2}\alpha_{--,-1/2}d^{--}_{-1/2}\nn
&&\hspace{1cm}+\alpha_{++,-3/2}\alpha_{++,-1/2}\alpha_{+-,-1/2}d^{-+}_{-1/2}-\frac{21}{8}\alpha_{++,-1/2}\alpha_{+-,-1/2}\alpha_{-+,-1/2}d^{--}_{-3/2}\nn
&&\hspace{1cm}+\frac{21}{8}\alpha_{++,-1/2}\alpha_{++,-1/2}\alpha_{--,-1/2}d^{--}_{-3/2}-\frac{29}{16}\alpha_{++,-3/2}d^{+-}_{-1/2}d^{-+}_{-1/2}d^{--}_{-1/2}\nn
&&\hspace{1cm}-\frac{21}{4}\alpha_{-+,-1/2}d^{--}_{-3/2}d^{+-}_{-1/2}d^{--}_{-1/2}-\frac{3}{16}\alpha_{++,-1/2}d^{--}_{-3/2}d^{+-}_{-1/2}d^{-+}_{-1/2}\nn
&&\hspace{1cm}-\frac{81}{32}\alpha_{++,-1/2}d^{--}_{-3/2}d^{++}_{-1/2}d^{--}_{-1/2}-\frac{81}{32}\alpha_{++,-1/2}d^{-+}_{-3/2}d^{+-}_{-1/2}d^{--}_{-1/2}\nn
&&\hspace{1cm}+\frac{75}{32}\alpha_{++,-1/2}d^{+-}_{-3/2}d^{-+}_{-1/2}d^{--}_{-1/2}\nn
&&\hspace{1cm}-\frac{21}{16}\alpha_{++,-1/2}\alpha_{+-,-1/2}\alpha_{-+,-1/2}d^{+-}_{-1/2}d^{-+}_{-1/2}d^{--}_{-1/2}\nn
&&\hspace{1cm}+\frac{21}{16}\alpha_{++,-1/2}\alpha_{++,-1/2}\alpha_{--,-1/2}d^{+-}_{-1/2}d^{-+}_{-1/2}d^{--}_{-1/2}\nn
&&\hspace{1cm}+\frac{21}{16}\alpha_{++,-1/2}\alpha_{+-,-1/2}\alpha_{+-,-1/2}\alpha_{-+,-1/2}\alpha_{-+,-1/2}d^{--}_{-1/2}\nn
&&\hspace{1cm}-\frac{21}{8}\alpha_{++,-1/2}\alpha_{++,-1/2}\alpha_{+-,-1/2}\alpha_{-+,-1/2}\alpha_{--,-1/2}d^{--}_{-1/2}\nn
&&\hspace{1cm}+\frac{21}{16}\alpha_{++,-1/2}\alpha_{++,-1/2}\alpha_{++,-1/2}\alpha_{--,-1/2}\alpha_{--,-1/2}d^{--}_{-1/2}
\Big]|0^{2-}_{R}\rangle
\eea
Because $\phi^{L}$ is a $\dot A$ charge singlet, $\phi^{L}_{-}$ is given by doing the following replacement in  $\phi^{L}_{+}$.
\be
\alpha_{A\pm,n}\rightarrow i\alpha_{A\mp,n}
\ee
Applying (\ref{lifting phi}) to the primary $\phi$, one finds the lift
\bea
E^{(2)}_{3,(1,0)}=\frac{64}{39}\pi^2\lambda^2
\eea

\subsection{Level-4}

The 15 primaries in the $\chi^{l}_{j=1,h}\bar \chi_{\bar j_{3}=-1}$ column can be organized into a $A$ charge triplet and $\dot A$ charge triplet $(1,1)$, a $A$ charge quintet and $\dot A$ charge singlet $(2,0)$ 
and a $A$ charge singlet and $\dot A$ charge singlet $(0,0)$. Their lifts are
\bea
E^{(2)}_{4,(1,1)}&=&3\pi^2\lambda^2\nn
E^{(2)}_{4,(2,0)}&=&\frac{5}{2}\pi^2\lambda^2\nn
E^{(2)}_{4,(0,0)}&=&\frac{2695}{1024}\pi^2\lambda^2
\eea
Since the states are very complicated, we will not write them explicitly here.

\subsection{Properties of lifting at $O(\lambda^2)$}\label{subsection unlifted}
In this subsection, we summarize the properties of the lift at $O(\lambda^2)$ for primaries in the theory with total winding $N=2$.

\b

(i) From the index, we expect that primaries in the columns $\chi^{l}_{j=1,h}\bar \chi_{\bar j_{3}=-1}$ and $\chi^{l}_{j=1,h}\bar \chi_{\bar j_{3}=0}$ are lifted. At order $O(\lambda^2)$, we find that primaries up to level-4 are all lifted. They are the states $\phi$ and $\phi_{+-}$ in the long multiplet (\ref{multiplets diagram}).

From eq. (\ref{right mover}) we see that the right movers of primaries in these two columns are  symmetric between the two copies. In the orbifold CFT the overall state of each string must be symmetric between the copies. 
Thus primaries in these two columns must be symmetric between the copies in both the left and  right sectors separately.

\b

(ii) From the index, we expect that primaries in the column $\chi^{l}_{j=1,h}\bar \chi_{\bar j_{3}=-1/2}$ in the $(1,1)$ sector are unlifted. At order $O(\lambda^2)$, we find that primaries at any level in this column are unlifted. The reason is the following. The right moving state is (\ref{r mover zero lifting}), which is antisymmetric between the two copies. Thus primaries in this column are antisymmetric in the left and right sectors separately. 
Because the projection operator (\ref{GN p s}) treats the left mover and right mover of the two copies symmetrically in the lifting computation, antisymmetry in both the left and right sectors leads to zero lift.

\b

(iii) The lifting properties of primaries in the $(2)$ sector determined by the details of operator (\ref{GN p s}). Because number of primaries in the $(2)$ sector grows faster than the numbers of primaries in the columns $\chi^{l}_{j=1,h}\bar \chi_{\bar j_{3}=-1}$ and $\chi^{l}_{j=1,h}\bar \chi_{\bar j_{3}=0}$ in the $(1,1)$ sector, most of the primaries in the $(2)$ sector will be unlifted. The lifted states are the $\phi_{+}$ and $\phi_{-}$ in the long multiplet (\ref{multiplets diagram}).

\section{Unlifted states in the untwisted sector}

In section \ref{sec lift}, we found that in our study upto level-4, the unlifted states in the $(1,1)$ sector  are  states that have  a right moving sector that is antisymmetric between the two copies. In this section, we will study the consequences of this antisymmetry in more detail, and generalize it to the situation where the component strings have higher twists.

 In section \ref{unlifted right 1,1}, we study the antisymmetry of the right movers for two singly wound strings. In section \ref{more general}, we generalize this antisymmetry to the situation where we have   many singly wound strings. In section \ref{strings M N}, we generalize this discussion to the case where we have two component strings with winding numbers $k_1$ and $k_2$.

\subsection{Right movers for the $(1,1)$ sector}\label{unlifted right 1,1}

Consider the case where $N=2$ and we have two singly wound strings. Let us focus on the right movers. We have used the fact that we can write the right movers on the two strings using the sum and difference of the oscillators on the two strings. Thus we get the global modes
\be\label{1 1 global}
\bar d_0^{+A(1)}+\bar d_0^{+A(2)}
\ee
which are symmetric between the two copies, and the antisymmetric modes
\be\label{1 1 anti}
\bar d_0^{+A(1)}-\bar d_0^{+A(2)}
\ee
which are antisymmetric between the two copies. We have seen that states where the right moving sector contains one antisymmetric mode are unlifted, while those containing zero or two antisymmetric modes are lifted. Thus unlifted states have an antisymmetric right sector while lifted states have a symmetric right sector. For example, the right moving states
\be\label{anti example 1}
\bar d^{+\pm(1)}_0\bar d^{+\pm(2)}_0
|\bar 0^-_R\rangle^{(1)}|\bar 0^-_R\rangle^{(2)}
=\frac{1}{2}(\bar d^{+\pm(1)}_0\bar d^{+\pm(2)}_0-\bar d^{+\pm(2)}_0\bar d^{+\pm(1)}_0)
|\bar 0^-_R\rangle^{(1)}|\bar 0^-_R\rangle^{(2)}
\ee
and 
\be
(\bar d^{++(1)}_0\bar d^{+-(2)}_0-\bar d^{++(2)}_0\bar d^{+-(1)}_0)
|\bar 0^-_R\rangle^{(1)}|\bar 0^-_R\rangle^{(2)}
\ee
are antisymmetric and thus lead to zero lift.

\subsection{More general states}\label{more general}

Let us now consider the case $N>2$. We will see that we can use the above notion of antisymmetrization of the right sector to make a large class of states that are unlifted at $O(\lambda^2)$. 

We first illustrate the idea by a simple example. Consider $3$ singly wound strings. Let the left and right sectors of each string be fermionic, so that each string is overall in a bosonic state.  Let the left sides of each string be in a different excited state:
\be
|\psi_L^1\rangle, ~~|\psi_L^2\rangle, ~~|\psi_L^3\rangle
\ee
Let the right sides be given by
\be
\bar d^{+A_1(1) }_0|\bar 0^-_R\rangle^{(1)}, ~~\bar d^{+A_2(2) }_0|\bar 0^-_R\rangle^{(2)}, ~~\bar d^{+A_3(3) }_0|\bar 0^-_R\rangle^{(3)}
\ee
Thus the overall state is
\be
|\Psi\rangle=-[\bar d^{+A_1(1) }_0\bar d^{+A_2(2) }_0\bar d^{+A_3(3) }_0]~|\psi_L^1\rangle|\bar 0^-_R\rangle^{(1)}\, |\psi_L^2\rangle|\bar 0^-_R\rangle^{(2)}\, |\psi_L^3\rangle|\bar 0^-_R\rangle^{(3)}
\ee
We now look at different cases

\b

(i) {All $A_i$ are the same:} ~Let us set $A_i=+$  for all $i$. Thus the state is
\be
|\Psi\rangle=-[\bar d^{++(1) }_0\bar d^{++(2) }_0\bar d^{++(3) }_0]~|\psi_L^1\rangle|\bar 0^-_R\rangle^{(1)}\, |\psi_L^2\rangle|\bar 0^-_R\rangle^{(2)}\, |\psi_L^3\rangle|\bar 0^-_R\rangle^{(3)}
\ee
As shown in (\ref{anti example 1}), the right mover is antisymmetric between any two copies.
If we twist any two strands together, then we get zero, so this state is unlifted to  $O(\lambda^2)$. 

\b

(ii) {One $A_i$ is different:}~Let us set $A_1=+, \, A_2=+,\, A_3=-$. The state is
\be
|\Psi_1\rangle
=
-[\bar d^{++(1) }_0\bar d^{++(2) }_0\bar d^{+-(3) }_0]~|\psi_L^1\rangle|\bar 0^-_R\rangle^{(1)}\, |\psi_L^2\rangle|\bar 0^-_R\rangle^{(2)}\, |\psi_L^3\rangle|\bar 0^-_R\rangle^{(3)}
\ee
There is no lift if we twist together strands $1,2$. But there is a lift for strands $2,3$. Thus let us take instead the antisymmetrized state
\bea
|\Psi_2\rangle&=&-[\bar d^{++(1) }_0\bar d^{++(2) }_0\bar d^{+-(3) }_0-\bar d^{++(1) }_0\bar d^{++(3) }_0\bar d^{+-(2) }_0-\bar d^{++(3) }_0\bar d^{++(2) }_0\bar d^{+-(1) }_0]\nn
&&~~~~~~~~~~~~~~~~~~~~~~~~~~~~|\psi_L^1\rangle|\bar 0^-_R\rangle^{(1)}\, |\psi_L^2\rangle|\bar 0^-_R\rangle^{(2)}\, |\psi_L^3\rangle|\bar 0^-_R\rangle^{(3)}\nn
&=&-[\bar d^{++(1) }_0\bar d^{++(2) }_0\bar d^{+-(3) }_0+\bar d^{++(1) }_0\bar d^{+-(2) }_0\bar d^{++(3) }_0+\bar d^{+-(1) }_0\bar d^{++(2) }_0\bar d^{++(3) }_0]\nn
&&~~~~~~~~~~~~~~~~~~~~~~~~~~~~|\psi_L^1\rangle|\bar 0^-_R\rangle^{(1)}\, |\psi_L^2\rangle|\bar 0^-_R\rangle^{(2)}\, |\psi_L^3\rangle|\bar 0^-_R\rangle^{(3)}
\eea
Now we find that there is no lift at $O(\lambda^2)$ from twisting any pair of strings.

We can now see that the above example is easily generalized. We can take any set of states with arbitrary left excitations, and antisymmetrize the right movers.  This will make the states unlifted at $O(\lambda^2)$. 
Note that when two strings have the same bosonic right movers, their antisymmetrization will lead to a vanishing of the state. Thus this method of making unlifted states can only be used when we have  fermionic right movers.

\subsection{Strings with winding numbers $k_1$ and $k_2$}
\label{strings M N}

In section \ref{unlifted right 1,1}, we had considered the case where two singly wound strings joined to a doubly wound string. In this subsection, we consider the analogous problem for the situation where a string with winding $k_1$ and a string with winding $k_2$ join to a string with winding $k_1+k_2$. 

Right movers in the $(k_1,k_2)$ sector can be built by the global modes
\be\label{M N global}
\bar d_0^{+A(k_1)}+\bar d_0^{+A(k_2)}
\ee
and the modes
\be\label{M N anti}
\frac{1}{k_1}\bar d_0^{+A(k_1)}-\frac{1}{k_2}\bar d_0^{+A(k_2)}
\ee
acting on the ground state
\be
|\bar 0^{k_1-}_R\rangle |\bar 0^{k_2-}_R\rangle
\ee
The global modes (\ref{M N global}) do not contribute to the lift. In  Appendix \ref{appmn}, we find that the operators $\bar G^{\bar \alpha (P)}_{\dot A,0}$ annihilate the right moving states containing one operator of type (\ref{M N anti}); the operators $\bar G^{\bar \alpha (P)}_{\dot A,0}$ do not annihilate the right moving states containing zero or two operators of (\ref{M N anti}).

Thus the contribution to the lift from the interaction joining together strings with winding numbers $k_1$ and $k_2$ is zero for states with right movers containing one operator of (\ref{M N anti}).

Note that the twist interaction can also break the strings of winding  $k_1$ or $k_2$  into shorter strings. But we do not consider this effect here; if the windings $k_1,k_2$ are much smaller than the total number $N$ of component strings, then combinatorical factors suppress the interactions where one strand of the string twists together with another strand of the same string.

The modes (\ref{M N global}) and (\ref{M N anti}) are the analogues of the symmetric and antisymmetric modes (\ref{1 1 global}) and (\ref{1 1 anti}). However, they do not have definite symmetry properties.

Now let us consider states with a fermionic right moving sector for each string.
Consider the right moving states
\bea\label{M N example 1}
&&\bar d^{+\pm(k_1)}_0\bar d^{+\pm(k_2)}_0
|\bar 0^{k_1-}_R\rangle|\bar 0^{k_2-}_R\rangle\nn
&=&-\frac{k_1k_2}{k_1+k_2}(\bar d^{+\pm(k_1)}_0+\bar d^{+\pm(k_2)}_0)(\frac{1}{k_1}\bar d^{+\pm(k_1)}_0-\frac{1}{k_2}\bar d^{+\pm(k_2)}_0)
|\bar 0^{k_1-}_R\rangle|\bar 0^{k_2-}_R\rangle
\eea
and 
\bea\label{M N example 2}
&&\bar d^{++(k_1)}_0\bar d^{+-(k_2)}_0-\bar d^{++(k_2)}_0\bar d^{+-(k_1)}_0\nn
&=&
-\frac{k_1k_2}{k_1+k_2}\Big[(\bar d^{++(k_1)}_0+\bar d^{++(k_2)}_0)(\frac{1}{k_1}\bar d^{+-(k_1)}_0-\frac{1}{k_2}\bar d^{+-(k_2)}_0)\nn
&&~~~~~~~~~~~~+(\bar d^{+-(k_1)}_0+\bar d^{+-(k_2)}_0)(\frac{1}{k_1}\bar d^{++(k_1)}_0-\frac{1}{k_2}\bar d^{++(k_2)}_0)\Big]
|\bar 0^{k_1-}_R\rangle|\bar 0^{k_2-}_R\rangle
\eea
States with right movers (\ref{M N example 1}) and (\ref{M N example 2}) will be unlifted because each term contains only one mode of (\ref{M N anti}). Note that these two right moving states are antisymmetric under the change $k_1\leftrightarrow k_2$. 
If we consider other possible right moving states that lead to zero lift, then we find that  there is no definite symmetric property of such states in general; what we have noted here is that in  the case where we have  fermionic right movers for each string the state is antisymmetric under the change $k_1\leftrightarrow k_2$.

Thus, we can generalize the discussion in subsection \ref{more general}. We take states with arbitrary winding numbers, arbitrary left excitations and only fermionic right movers.  Antisymmetrizing the fermionic right movers will make the states unlifted at $O(\lambda^2)$, in the approximation  where we only consider the effect of twisting two strings into a longer string and ignore the breaking of the strings to shorter strings.

\section{Discussion}

To understand the AdS/CFT correspondence we need to match quantities between the gravity and field theory descriptions. The principal difficulty in this match is the fact that one of these descriptions is weakly coupled, then the other is strongly coupled. There have been much progress  in understanding black holes  by comparing quantities in the free CFT to results with  weakly coupled gravity; however a more detailed comparison will necessarily need us to perturb the CFT away from its free (i.e. orbifold) point. The most basic quantity we can study under such a perturbation is the energy levels of states in the CFT as a function of the perturbation parameter $\lambda$.

In this paper we have considered states which are BPS at the orbifold point, and asked which states are lifted and by how much as we perturb away from this point. For most of this paper we worked with the simplest system which allows a computation of the lift: the case $N=2$ where we have $2$ copies of the basic $c=6$ CFT making up the orbifold theory. The lifted states must join into a supermultiplet in which all members will have the same lift; we found this multiplet structure and the value of the lift to order $O(\lambda^2)$ for the first $4$ levels above the Ramond ground state. For the first $3$ levels we also explicitly write down the wavefunctions of the lifted states. We found that all the states that are allowed to lift by the index are in fact lifted at $O(\lambda^2)$.

We observed that the unlifted states in the untwisted sector have a right moving sector which was antisymmetric between the two copies of the $c=6$ CFT. We then extended this observation to the case  $N>2$, where we found states which would remain unlifted to order $O(\lambda^2)$ by using a feature similar to this antisymmetry. 

For large $N$, one can consider the gravity dual. In this gravity description a large class of supergravity states called `superstrata' were found in  \cite{Bena:2015bea,Bena:2016agb,Bena:2016ypk,Bena:2017xbt,Ceplak:2018pws,Heidmann:2019zws}. These are extremal states that describe fully backreacted solutions arising from massless string quanta placed in the $AdS_3\times S^3\times T^4$ geometry. 
The analysis of lifting  carried out in the present paper may help us to understand the nature of the CFT states that are dual to these BPS gravity solutions.
We hope to return to this issue in a future work.

\section*{Acknowledgements}

We would like to thank Nathan Benjamin, Stefano Giusto, Shaun Hampton, Rodolfo Russo, David Turton,  Ida Zadeh and  Xinan Zhou, for many helpful discussions.  This work is supported in part by DOE grant de-sc0011726. 

\appendix

\section{Contracted large $\mathcal N=4$ superconformal algebra}\label{commutators}
We follow the notation in the appendix of \cite{hmz}. The indices $\alpha=(+,-)$ and $\bar \alpha=(+,-)$ correspond to the subgroups $SU(2)_L$ and $SU(2)_R$ arising from rotations on $S^3$. The indices $ A=(+,-)$ and $\dot A=(+,-)$ correspond to the subgroups $SU(2)_1$ and $SU(2)_2$ arising  from rotations in $T^4$. We use the convention
\be
\epsilon_{+-}=1, ~~~\epsilon^{+-}=-1
\ee
The commutation relations for the contracted large $\mathcal N=4$ superconformal algebra are
\bea\label{app com a d}
[\alpha_{A\dot{A},m},\alpha_{B\dot{B},n}] &=& -\frac{c}{6}m\epsilon_{A B}\epsilon_{\dot A \dot{B}}\delta_{m+n,0}\cr
\{d^{\alpha A}_r , d^{\beta B}_s\}  &=&-\frac{c}{6}\epsilon^{\alpha\beta}\epsilon^{AB}\delta_{r+s,0}
\eea
\bea\label{app com current a d}
[L_m,\alpha_{A\dot{A},n}] &=&-n\alpha_{A\dot{A},m+n} ~~~~~~~[L_m ,d^{\alpha A}_r] =-({m\over2}+r)d^{\alpha A}_{m+r}\cr
\lbrace G^{\alpha}_{\dot{A},r} ,  d^{\beta B}_{s} \rbrace&=&i\epsilon^{\alpha\beta}\epsilon^{AB}\alpha_{A\dot{A},r+s}~~~~~~~
[G^{\alpha}_{\dot{A},r} , \alpha_{B \dot{B},m}]=  -im\epsilon_{AB}\epsilon_{\dot{A}\dot{B}}d^{\alpha A}_{r+m}\cr
[J^a_m,d^{\alpha A}_r] &=&{1\over 2}(\sigma^{Ta})^{\alpha}_{\beta}d^{\beta A}_{m+r}
\eea
\bea\label{app com currents}
[L_m,L_n] &=& {c\over12}m(m^2-1)\delta_{m+n,0}+ (m-n)L_{m+n}\cr
[J^a_{m},J^b_{n}] &=&{c\over12}m\delta^{ab}\delta_{m+n,0} +  i\epsilon^{ab}_{\,\,\,\,c}J^c_{m+n}\cr
\lbrace G^{\alpha}_{\dot{A},r} , G^{\beta}_{\dot{B},s} \rbrace&=&  \epsilon_{\dot{A}\dot{B}}\bigg[\epsilon^{\alpha\beta}{c\over6}(r^2-{1\over4})\delta_{r+s,0}  + (\sigma^{aT})^{\alpha}_{\gamma}\epsilon^{\gamma\beta}(r-s)J^a_{r+s}  + \epsilon^{\alpha\beta}L_{r+s}  \bigg]\cr
[J^a_{m},G^{\alpha}_{\dot{A},r}] &=&{1\over2}(\sigma^{aT})^{\alpha}_{\beta} G^{\beta}_{\dot{A},m+r}\cr
[L_{m},J^a_n]&=& -nJ^a_{m+n}\cr
[L_{m},G^{\alpha}_{\dot{A},r}] &=& ({m\over2}  -r)G^{\alpha}_{\dot{A},m+r}
\eea
We define $J^{+}_n, J^-_n$ as
\bea
J^+_n &=& J^1_n + i J^2_n\cr
J^-_n&=& J^1_n - i J^2_n
\eea
From (\ref{app com current a d}), one can see that $d^{\alpha A}_{n}$ with $\alpha=+,-$ is a $SU(2)_L$ charge doublet. We have
\bea
[J^{+}_m,d^{+ A}_r] &=& 0,\qquad~~~~~ [J^{-}_m,d^{+ A}_r] ~=~ d^{-A}_{m+r}\cr
[J^{+}_m,d^{- A}_r] &=& d^{+A}_{m+r},\qquad [J^{-}_m,d^{- A}_r] ~=~ 0
\eea
From (\ref{app com currents}), one can see that $G^{\alpha}_{\dot{A},r}$  with $\alpha=+,-$ is also a $SU(2)_L$ charge doublet. We have
\bea
[J^{+}_{m},G^{+}_{\dot{A},r}]  &=& 0 ,\qquad\qquad ~~~[J^{-}_{m},G^{+}_{\dot{A},r}]  ~=~ G^{-}_{\dot{A},m+r}\cr
[J^{+}_{m},G^{-}_{\dot{A},r}]  &=&G^{+}_{\dot{A},m+r},\qquad ~[J^{-}_{m},G^{-}_{\dot{A},r}]  ~=~ 0 
\eea
It is believed that the contracted large $\mathcal N=4$ superconformal algebra is an exact symmetry at any point of the moduli space.

Now let's consider the orbifold point.
Look at the winding sector $(k_1,k_2,...,k_i,...)$
with the total winding $N=\sum_{i} k_i$. For the $i$th twisted set of copies  with winding number $k_{i}$,
we have following mode expansions on the cylinder.
\bea\label{i string modes}
\alpha^{(i)}_{A \dot A,n}=\frac{1}{2\pi}\int_{\sigma=0}^{2\pi k_{i}}\p_{w}X^{(i)}_{A \dot A}(w)e^{nw}dw\nn
d^{\alpha A (i)}_{n}=\frac{1}{2\pi i}\int_{\sigma=0}^{2\pi k_{i}}\psi^{\alpha A(i)}(w)e^{nw}dw
\eea
In terms of $\alpha$ and $d$ modes, the $J$, $G$ and $L$ modes can be written as
\bea\label{i string modes J G L}
J^{a(i)}_m &=& {1\over 4 k_{i}}\sum_{r}\epsilon_{AB}d^ {\gamma B(i)}_r\epsilon_{\alpha\gamma}(\sigma^{aT})^{\alpha}_{\beta}d^ {\beta A(i)}_{m-r},\qquad a=1,2,3\cr
J^{3(i)}_m &=&  - {1\over 2 k_{i}}\sum_{r} d^ {+ +(i)}_{r}d^ {- -(i)}_{m-r} - {1\over 2 k_{i}}\sum_{r}d^ {- +(i)}_r d^ {+ -(i)}_{m-r}\cr
J^{+(i)}_m&=&\frac{1}{k_{i}}\sum_{r}d^ {+ +(i)}_rd^ {+ -(i)}_{m-r} ,\qquad J^{-(i)}_m=\frac{1}{k_{i}}\sum_{r}d^ {--(i)}_rd^ {- +(i)}_{m-r}\cr
G^{\alpha(i)}_{\dot{A},r} &=& -\frac{i}{k_{i}}\sum_{n}d^ {\alpha A(i)}_{r-n} \alpha^{(i)}_{A\dot{A},n}\cr
L^{(i)}_m&=& -{1\over 2 k_i}\sum_{n} \epsilon^{AB}\epsilon^{\dot A \dot B}\alpha^{(i)}_{A\dot{A},n}\alpha^{(i)}_{B\dot{B},m-n}- {1\over 2 k_i}\sum_{r}(m-r+{1\over2})\epsilon_{\alpha\beta}\epsilon_{AB}d^ {\alpha A(i)}_r d^ {\beta B(i)}_{m-r}
\eea
Let $q$ be an integer. 
The mode numbers for $\alpha,L,J$ are $n=q/k_{i}$.
In the R sector, the mode numbers for $d$ and $G$ are $n=q/k_{i}$.
In the NS sector, the mode numbers for $d$ and $G$ are $n=(q+\frac{1}{2})/k_{i}$.
The modes (\ref{i string modes})  and (\ref{i string modes J G L}) satisfy the contracted large $\mathcal N=4$ superconformal algebra (\ref{app com a d})(\ref{app com current a d})(\ref{app com currents}) with $c=6k_{i}$. 

We define the global modes $O^{(g)}_n$ by summing the terms from each copy
\be\label{global modes app}
O^{(g)}_n=\sum_{i}O^{(i)}_n
\ee
where the modes $O$ can be modes of $\alpha,~d,~L,~J,~G$.
The global modes satisfy the contracted large $\mathcal N=4$ superconformal algebra (\ref{app com a d})(\ref{app com current a d})(\ref{app com currents}) with $c=6N$. It is believed that global modes satisfy the algebra at any point in the moduli space.

\section{The $\mathcal N=4$ character}\label{character}

In this appendix, we will present the character of the small and the contracted large $\mathcal N=4$ superconformal algebra. The characters are defined as follows:
\be
\chi_{j,h}(q,y)={\rm Tr}(-1)^{2 J^{3}_{0}}q^{L_{0}-\frac{c}{24}}y^{2 J^{3}_0}
\ee
where the trace is over states in the irreducible representation. In this paper we will work in the Ramond sector. The primaries in these represenations have dimension $h$, which is the eigenvalue of $L_{0}-\frac{c}{24}$, and SU(2) charge $j$.

\subsection{The small $\mathcal N=4$ character}

The characters were computed in \cite{Eguchi:1987sm,Eguchi:1987wf}.
The central charge is given by $c=6m$, where $m$ is an integer. We use $\chi^{(S)}$ to label the character in the small $\mathcal N=4$ algebra.
There are $m+1$ short representations with $h=0$ and $j=0,1/2,\dots,m/2$. Their characters are
\bea\label{small chara short}
\chi^{s(S)}_{j;m}(\tau,z)&=&
(-1)^{2j} \left(\frac{i \theta_{1}(\tau,z)^2}{\theta_{1}(\tau,2 z)\eta(\tau)^3}\right)\nn
&&~~~~~~\sum_{k\in\mathbb{Z}}\frac{q^{(m+1)k^2+k}y^{2(m+1)k+1}}{(1-yq^k)^2}
\left(q^{k(2j+1)}y^{2j+1}-q^{-k(2j+1)}y^{-(2j+1)}\right)
\eea
where we define $q=e^{2\pi i \tau}$ and $y=e^{2\pi i z}$.
There are $m$ long representations with $h>0$ and $j=1/2,1,\dots,m/2$. Their characters are
\be\label{small chara long}
\chi^{l(S)}_{j,h;m}(\tau,z)=q^{h}(-1)^{2j} \left(\frac{i \theta_{1}(\tau,z)^2}{\theta_{1}(\tau,2 z)\eta(\tau)^3}\right)\sum_{k\in\mathbb{Z}}q^{(m+1)k^2}y^{2(m+1)k}
\left(q^{2kj}y^{2j}-q^{-2kj}y^{-2j}\right)
\ee

\subsection{The contracted large $\mathcal N=4$ character}

The characters were computed in \cite{Petersen:1989zz,Petersen:1989pp}.
The central charge is given by $c=6m$, where $m$ is an integer. 
There are $m$ short representations with $h=0$ and $j=1/2,1,\dots,m/2$
\be
\chi^{s}_{j;m}(\tau,z)=\frac{q^{1/8}L(\tau,z)}{\eta(\tau)^3}\chi^{s(S)}_{j-1/2;m-1}(\tau,z)
\ee
There are $m-1$ long representations with $h>0$ and $j=1,\dots,m/2$
\be
\chi^{l}_{j,h;m}(\tau,z)=\frac{q^{1/8}L(\tau,z)}{\eta(\tau)^3}\chi^{l(S)}_{j-1/2,h;m-1}(\tau,z)
\ee
where
\be
L(\tau,z)=\chi^{l(S)}_{j=1/2,h=0;m=1}(\tau,z)
\ee
Here $\chi^{s(S)}$ and $\chi^{l(S)}$ are characters of the short and long multiplets (\ref{small chara short})(\ref{small chara long}) of the small $\mathcal N=4$ algebra.

\section{The effect of the twist operator}\label{s twist}

The operator (\ref{GN p s}) contains the twist operator $\sigma_2$. The action of this twist was studied in \cite{Avery:2010er,Avery:2010hs}. Here we recall some results about this action which will be of use to us later in the computation of $E^{(2)}$.

We consider only the left sector. Start in the twist sector $N=(1,1)$ where we have two singly wound copies of the CFT. Let the initial state be the Ramond ground state $|0^{-}_{R}\rangle|0^{-}_{R}\rangle$. Let us apply the twist operator $\sigma^+_2$ at the position $w_0$ on the cylinder. This action generates the state $|\chi\rangle$
\bea\label{chi}
|\chi\rangle=\sigma^{+}_{2}(w_{0})|0^{-}_{R}\rangle|0^{-}_{R}\rangle&=&e^{\sum_{m\geq 1/2,n\geq 1/2}\gamma^{B}_{mn}[-\alpha_{++,-m}\alpha_{--,-n}+\alpha_{-+,-m}\alpha_{+-,-n}]}\nn
&&e^{\sum_{m\geq 1/2,n\geq 1/2}\gamma^{F}_{mn}[d^{++}_{-m}d^{--}_{-n}-d^{+-}_{-m}d^{-+}_{-n}]}
|0^{2-}_R\rangle
\eea
where  
\bea
\gamma^{B}_{m'+1/2,n'+1/2}&=&\frac{2}{(2m'+1)(2n'+1)}\frac{a^{2(m'+n'+1)}\Gamma[\frac{3}{2}+m']\Gamma[\frac{3}{2}+n']}{(1+m'+n')\pi \Gamma[m'+1]\Gamma[n'+1]}\nn
\gamma^{F}_{m'+1/2,n'+1/2}&=&-\frac{a^{2(m'+n'+1)}\Gamma[\frac{3}{2}+m']\Gamma[\frac{3}{2}+n']}{(2n'+1)\pi(1+m'+n') \Gamma[m'+1]\Gamma[n'+1]}
\eea
where $a=e^{w_{0}/2}$ and $m',n'$ are negative integers.

For states containing one oscillator excitation on the vacuum $|0^{-}_{R}\rangle|0^{-}_{R}\rangle$, we have
\bea\label{a A}
&&\sigma^{+}_{2}(w_{0})\alpha^{(\mathcal A)}_{B\dot B,n} |0^{-}_{R}\rangle|0^{-}_{R}\rangle\nn
&=&2\sum_{p'\leq -1}\frac{i}{\pi}\frac{\Gamma[\frac{1}{2}-n]}{\Gamma[-n]}
\frac{\Gamma[-\frac{1}{2}-p']}{\Gamma[-p']}\frac{a^{2(n-p')-1}}{2n-2p'-1}
\alpha_{B\dot B,p'+1/2}|\chi\rangle
\eea
\bea\label{d+ A}
&&\sigma^{+}_{2}(w_{0})d^{+B(\mathcal A)}_{n}|0^{-}_{R}\rangle|0^{-}_{R}\rangle\nn
&=&2\sum_{p'\leq -1}\frac{i}{\pi}\frac{\Gamma[\frac{1}{2}-n]}{\Gamma[1-n]}
\frac{\Gamma[\frac{1}{2}-p']}{\Gamma[-p']}\frac{a^{2(n-p')-1}}{2n-2p'-1}
d^{+B}_{p'+1/2}|\chi\rangle
\eea
and
\bea\label{d- A}
&&\sigma^{+}_{2}(w_{0})d^{-B(\mathcal A)}_{n}|0^{-}_{R}\rangle|0^{-}_{R}\rangle\nn
&=&2\sum_{p'\leq -1}\frac{i}{\pi}\frac{\Gamma[\frac{1}{2}-n]}{\Gamma[-n]}
\frac{\Gamma[-\frac{1}{2}-p']}{\Gamma[-p']}\frac{a^{2(n-p')-1}}{2n-2p'-1}
d^{-B}_{p'+1/2}|\chi\rangle
\eea
where $\alpha^{(\mathcal A)}_{B\dot B,n}$ and $d^{\beta B(\mathcal A)}_{n}$ are the antisymmetric operators defined in (\ref{def anti}). We use these antisymmetric operators to build primaries in section \ref{primaries left}.

We can also start with initial states that have more than one oscillator excitation; i.e., we can compute
\be
\sigma^{+}_{2}(w_{0})\Big(\prod_{i}O_{i,-n_{i}}\Big)|0^{-}_{R}\rangle|0^{-}_{R}\rangle
\ee
The general method to compute the final state in this situation was given in \cite{Avery:2010hs}. We first do all possible contractions in $\prod_{i}O_{i,-n_{i}}$ using the following rules
\bea
&&C\Big[\alpha^{(\mathcal A)}_{A\dot A,m}\alpha^{(\mathcal A)}_{B\dot B,n}\Big]=
2\epsilon_{AB}\epsilon_{\dot A \dot B}(2ia)^{2 m+2 n}\nn
&&\Big[
\sum_{k= 0}^{-n-1} 
{}^{m}C_{-m-n-k}{}^{n}C_{k}
(n+k)
-\sum_{k= 0}^{-m-1}\sum_{q= 0}^{-n-1} {}^{m}C_{k} {}^{n}C_{q} {}^{m+k}C_{-(n+q)}
(-1)^{m-k+1}(n+q)
\Big]\nn
\eea
and
\bea
&&C\Big[d^{(\mathcal A)+ A}_m d^{(\mathcal A)- B}_n\Big]\nn
&&=2\epsilon^{AB}(2ia)^{2m+2n}\Big[\sum_{q=0}^{-n-1}~{}^{m-1}C_{-m-n-q}~{}^{n}C_{q}
+2\sum_{q=0}^{-n-1}~{}^{m-1}C_{-m-n-q-1}~{}^{n}C_{q}\nn
&&\hspace{4cm}+\sum_{k=0}^{-m}\sum_{q=0}^{-n-1}(-1)^{m-k}~~
{}^{m-1}C_k~{}^{n}C_q~{}^{m+k-1}C_{-n-q-1}\nn
&&\hspace{4cm}+2\sum_{k=0}^{-m-1}\sum_{q=0}^{-n-1}(-1)^{m-k-1}~~
{}^{m-1}C_k~{}^{n}C_q~{}^{m+k}C_{-n-q-1}
\Big]
\eea
Then each of oscillators left is moved separately to the final state as indicated in the relations (\ref{a A}-\ref{d- A}) discussed above where we had only one initial operator.

\section{Twisting strings with winding numbers $k_1$ and $k_2$}\label{appmn}
 
The right moving part of the   operators $\bar G^{\bar \alpha (P)}_{\dot A,0}$ is given by
$\mathcal P \bar \sigma^{\bar \alpha}$. 
The effect of twist operator $\bar \sigma^{\bar \alpha}$ on strings with winding numbers $k_1$ and $k_2$ was found in \cite{Carson:2014ena}. Including the projection operator $\mathcal P$, the effect of this operator on the ground state is
\be
\mathcal P \bar \sigma_2^-|\bar 0^{k_1-}_R\rangle |\bar 0^{k_2-}_R\rangle
=0
\ee
\be\label{sigma+ on ground}
\mathcal P \bar \sigma_2^+|\bar 0^{k_1-}_R\rangle |\bar 0^{k_2-}_R\rangle
=
C_{k_1,k_2}^{1/2}
|\bar 0^{(k_1+k_2)-}_R\rangle
\ee
where
\be
C_{k_1,k_2}=\frac{k_1+k_2}{2k_1 k_2}
\ee
A fermionic mode applied to one of the strings before the twist gives rise to a fermionic mode on the joined string obtained after the twist
\be\label{d on M}
\mathcal P \bar \sigma_{2}^{+}
\bar d^{+A(k_1)}_0|\bar 0^{k_1-}_R\rangle |\bar 0^{k_2-}_R\rangle={k_1\over k_1+k_2} C_{k_1,k_2}^{1/2}\bar d^{+A(k_1+k_2)}_0|\bar 0^{(k_1+k_2)-}_R\rangle
\ee
\be\label{d on N}
\mathcal P \bar\sigma_{2}^{+}
\bar d^{+A(k_2)}_0|\bar 0^{k_1-}_R\rangle |\bar 0^{k_2-}_R\rangle={k_2\over k_1+k_2} C_{k_1,k_2}^{1/2} \bar d^{+A(k_1+k_2)}_0|\bar 0^{(k_1+k_2)-}_R\rangle
\ee
Thus
\be\label{sigma+ kill d}
\mathcal P \bar \sigma_{2}^{+}\left ( {1\over k_1} \bar d^{+A(k_1)}_0-{1\over k_2}\bar d^{+A(k_2)}_0 \right ) |\bar 0^{k_1-}_R\rangle |\bar 0^{k_2-}_R\rangle= 0
\ee
We can also consider the action of $\mathcal P \bar\sigma^-$, but in this case we see that the final state on the $k_1+k_2$ wound string must have charge $\bar\alpha=-1$, which is not possible for a Ramond ground state. Thus we have
\be\label{sigma- kill d}
\mathcal P \bar \sigma_{2}^{-}\left ( {1\over k_1} \bar d^{+A(k_1)}_0-{1\over k_2}\bar d^{+A(k_2)}_0 \right ) |\bar 0^{k_1-}_R\rangle |\bar 0^{k_2-}_R\rangle= 0
\ee

For the case where we have two $\bar d_0$ operators on the initial strings, each such operator can be moved to the final state string using (\ref{d on M}) and (\ref{d on N}).  We thus have
\be\label{sigam + on dd}
\mathcal P \bar \sigma^{+}_2 (\frac{1}{k_1}\bar d_0^{++(k_1)}-\frac{1}{k_2}\bar d_0^{++(k_2)})(\frac{1}{k_1}\bar d_0^{+-(k_1)}-\frac{1}{k_2}\bar d_0^{+-(k_2)})|\bar 0^{k_1-}_R\rangle |\bar 0^{k_2-}_R\rangle=0
\ee
In general there is also a contraction term between the two initial state operators. However this contraction term will vanish in our present situation, since both the $\bar d_0$ operators have positive $\bar\alpha$ charge, and so they cannot contract with each other.

Now consider the action of $\mathcal P \bar\sigma^-$ on the state with two initial excitations
\bea\label{sigma- on dd}
&&\mathcal P \bar \sigma^{-}_2 (\frac{1}{k_1}\bar d_0^{++(k_1)}-\frac{1}{k_2}\bar d_0^{++(k_2)})(\frac{1}{k_1}\bar d_0^{+-(k_1)}-\frac{1}{k_2}\bar d_0^{+-(k_2)})|\bar 0^{k_1-}_R\rangle |\bar 0^{k_2-}_R\rangle \nn
&=& [\bar J^-_0, \mathcal P \bar \sigma^{+}_2] (\frac{1}{k_1}\bar d_0^{++(k_1)}-\frac{1}{k_2}\bar d_0^{++(k_2)})(\frac{1}{k_1}\bar d_0^{+-(k_1)}-\frac{1}{k_2}\bar d_0^{+-(k_2)})|\bar 0^{k_1-}_R\rangle |\bar 0^{k_2-}_R\rangle\nn
&=& -\mathcal P \bar \sigma^{+}_2 \bar J^-_0  (\frac{1}{k_1}\bar d_0^{++(k_1)}-\frac{1}{k_2}\bar d_0^{++(k_2)})(\frac{1}{k_1}\bar d_0^{+-(k_1)}-\frac{1}{k_2}\bar d_0^{+-(k_2)})|\bar 0^{k_1-}_R\rangle |\bar 0^{k_2-}_R\rangle\nn
&=& -\mathcal P \bar \sigma^{+}_2   (\frac{1}{k_1}\bar d_0^{-+(k_1)}-\frac{1}{k_2}\bar d_0^{-+(k_2)})(\frac{1}{k_1}\bar d_0^{+-(k_1)}-\frac{1}{k_2}\bar d_0^{+-(k_2)})|\bar 0^{k_1-}_R\rangle |\bar 0^{k_2-}_R\rangle\nn
&=& -\mathcal P \bar \sigma^{+}_2
(\frac{1}{k_1}+\frac{1}{k_2})|\bar 0^{k_1-}_R\rangle |\bar 0^{k_2-}_R\rangle\nn
&=& -\frac{k_1+k_2}{ k_1 k_2}C_{k_1,k_2}^{1/2}|\bar 0^{(k_1+k_2)-}_R\rangle
\eea
where to get the third line we used (\ref{sigam + on dd}) and to get the fifth line we used the commutator
(\ref{app com a d}).

In summary, we find that 
the operators $\bar G^{\bar \alpha (P)}_{\dot A,0}$  annihilate the right moving states which contain one  operator of the type (\ref{M N anti}); this is shown in   (\ref{sigma+ kill d}) and (\ref{sigma- kill d}). But the operators $\bar G^{\bar \alpha (P)}_{\dot A,0}$  do not annihilate the right moving sector containing zero or two operators of the type (\ref{M N anti}); this is shown by   (\ref{sigma+ on ground}) and (\ref{sigma- on dd}).

\end{document}